\documentclass[11pt]{article}%

\usepackage{amsmath}
\usepackage{amsfonts}
\usepackage{amssymb}
\usepackage{graphicx}
\usepackage{epsfig}
\begin{document}
\begin{center}
{\bf {\large On the $SU(3)$ Parametrization of
Qutrits}\footnote{Paper presented at
The 12th Central European Workshop on Quantum Optics, 6-9 June 2005, Ankara, Turkey.}\\

\bigskip

A. T. B\"{o}l\"{u}kba\c{s}{\i}\footnote{E.mail:
abolukbasi@ku.edu.tr} ,
 \underline{T. Dereli}\footnote{E.mail: tdereli@ku.edu.tr}}

\medskip

Department of Physics, Ko\c{c} University\\
34450 Sar{\i}yer, \.{I}stanbul, Turkey\\
\bigskip
{\bf Abstract}
\end{center}
\noindent {\small  Parametrization of qutrits on the complex
projective plane $\mathcal{C}P^{2}=SU(3)/U(2)$ is given
explicitly. A set of constraints that characterize mixed state
density matrices is found.}

\vskip 8mm

\noindent Many recent ideas of quantum information theory are
based on the notion of qubits. A qubit may be represented by a
point on the Poincar\'{e} sphere $S^2$ that is homeomorhic to the
complex projective line $ \mathcal{H}^{(2)}= \mathcal{C}P^{1} =
SU(2)/U(1)$. A similar parametrization in the case of higher
dimensional quantum systems is desirable both from theoretical [1]
and technical points of view [2], [3]. A qutrit may be represented
by a point on the complex projective plane
$\mathcal{H}^{(3)}=\mathcal{C}P^{2}=SU(3)/U(2)$ . Such a
representation is given explicitly in terms of Gell-Mann matrices
[4]. We determine a set of constraints that characterize mixed
states of qutrits below.

A qutrit $|\psi>=\alpha_{0}|0>+\alpha_{1}|1>+\alpha_{2}%
|2>$where $\alpha_{0},\alpha_{1},\alpha_{2}\in\mathbf{{C}}$ , $|\alpha
_{0}|^{2}+|\alpha_{1}|^{2}+|\alpha_{2}|^{2}=1$ , is a state vector in the
Hilbert space of states $\mathcal{H}^{(3)}$ of a 3-level system. It is spanned
by an orthonormal basis $\{|0>,|1>,|2>\}$ which in matrix notation reads
\[
|0>\rightarrow\left(
\begin{array}
[c]{c}%
1\\
0\\
0
\end{array}
\right)  ,|1>\rightarrow\left(
\begin{array}
[c]{c}%
0\\
1\\
0
\end{array}
\right)  ,|2>\rightarrow\left(
\begin{array}
[c]{c}%
0\\
0\\
1
\end{array}
\right)  .
\]
Therefore
\[
|\psi>\rightarrow\left(
\begin{array}
[c]{c}%
\alpha_{0}\\
\alpha_{1}\\
\alpha_{2}%
\end{array}
\right)  \in\mathbf{C}^{3}\simeq\mathbf{R}^{6}.
\]
Since $|\alpha_{0}|^{2}+|\alpha_{1}|^{2}+|\alpha_{2}|^{2}=1$ and since
$|\psi>$ is determined up to a multiplicative phase factor, dim$\mathcal{H}%
^{(3)}=4$. \newline Any $3\times3$ density matrix can be written
as
\[
\rho=\frac{1}{3}(I+\sqrt{3}\vec{n}\cdot\vec{\lambda})
\]
where $\vec{n}$ is a real 8-vector, and components of $\vec{\lambda}$ are the
(Hermitian, traceless) Gell-Mann matrices
\[
\lambda_{1}=\left(
\begin{array}
[c]{ccc}%
0 & 1 & 0\\
1 & 0 & 0\\
0 & 0 & 0
\end{array}
\right) ,  \text{ \ \ }\lambda_{2}=\left(
\begin{array}
[c]{ccc}%
0 & -i & 0\\
i & 0 & 0\\
0 & 0 & 0
\end{array}
\right) ,  \text{ \ }\lambda_{4}=\left(
\begin{array}
[c]{ccc}%
0 & 0 & 1\\
0 & 0 & 0\\
1 & 0 & 0
\end{array}
\right) ,  \text{\ }%
\]%
\[
\text{ \ \ }\lambda_{5}=\left(
\begin{array}
[c]{ccc}%
0 & 0 & -i\\
0 & 0 & 0\\
i & 0 & 0
\end{array}
\right) , \text{ \ }\lambda_{6}=\left(
\begin{array}
[c]{ccc}%
0 & 0 & 0\\
0 & 0 & 1\\
0 & 1 & 0
\end{array}
\right) ,  \text{\ }\lambda_{7}=\left(
\begin{array}
[c]{ccc}%
0 & 0 & 0\\
0 & 0 & -i\\
0 & i & 0
\end{array}
\right) , \text{ }%
\]%
\[
\lambda_{3}=\left(
\begin{array}
[c]{ccc}%
1 & 0 & 0\\
0 & -1 & 0\\
0 & 0 & 0
\end{array}
\right) , \text{ \ \ }\lambda_{8}=\frac{1}{\sqrt{3}}\left(
\begin{array}
[c]{ccc}%
1 & 0 & 0\\
0 & 1 & 0\\
0 & 0 & -2
\end{array}
\right) .
\]
The product of two Gell-Mann matrices is given by
\[
\lambda_{j}\lambda_{k}=\frac{2}{3}\delta_{jk}+\sum_{l}d_{jkl}\lambda_{l}%
+i\sum_{l}f_{jkl}\lambda_{l}%
\]
where $j,k=1,2,\dots,8$. The $f$-symbols (structure constants of
the Lie algebra $\textsf{su}(3)$) are totally anti-symmetric :
\begin{align}
f_{123}  &  =1,f_{458}=f_{678}=\frac{\sqrt{3}}{2},\nonumber\\
f_{147}  &  =f_{246}=f_{257}=f_{345}=f_{516}=f_{637}=\frac{1}{2},\nonumber
\end{align}
and the $d$-symbols are totally symmetric:
\begin{align}
d_{118}  &  =d_{228}=d_{338}=-d_{888}=\frac{1}{\sqrt{3}},\text{
\ \ \ \ \ \ \ \ }d_{448}=d_{558}=d_{668}=d_{778}=-\frac{1}{2\sqrt{3}%
},\nonumber\\
d_{146}  &  =d_{157}=-d_{247}=d_{256}=d_{344}=d_{355}=-d_{366}=-d_{377}%
=\frac{1}{2}.\nonumber
\end{align}

\noindent Given two real 8-vectors $\vec{a}$ and $\vec{b}$,  we
define their  Euclidean inner  product
\[
\vec{a} \cdot\vec{b}= \sum_{k} a_{k}b_{k} \quad ,
\]
skew-symmetric vector $\wedge$-product
\[
(\vec{a} \wedge\vec{b})_{j}= \sqrt{3} \sum_{k,l}
f_{jkl}~a_{k}b_{l} \quad ,
\]
and symmetric vector $\star$-product
\[
(\vec{a}\star\vec{b})_{j}=\sqrt{3}\sum_{k,l}d_{jkl}~a_{k}b_{l}
\quad .
\]
The pure states that satisfy $\rho^{2}=\rho$  are therefore
characterized by
\[
|\vec{n}|^{2}=1 \quad \text{ \ and } \quad \vec{n}\star\vec{n}=\vec{n} \quad \text{.}%
\]

Suppose that
$\rho=\frac{1}{3}(I+\sqrt{3}\vec{n}\cdot\vec{\lambda})$ \ is the
density matrix of a mixed state. It is Hermitian, positive with
trace equal to $1$. Therefore all the eigenvalues
$x_{1},x_{2},x_{3}$ are positive and add to one:
$x_{1}+x_{2}+x_{3} = 1$.  The Cayley-Hamilton equation satisfied
by $\rho$ reads
\[
\rho^{3}-\rho^{2}+(x_{1}x_{2}+x_{2}x_{3}+x_{1}x_{3})\rho-x_{1}x_{2}x_{3}I
= 0 \quad .
\]
The following inequalities hold:
\[
\frac{1}{3}\geq x_{1}x_{2}+x_{2}x_{3}+x_{1}x_{3}\geq 0\quad , \quad\frac{1}%
{27}\geq x_{1}x_{2}x_{3}\geq 0 \quad .
\]
Starting from these, a straightforward computation shows that the
necessary and sufficient conditions for $\rho =
\frac{1}{3}(I+\sqrt{3}n\cdot\lambda)$ to be a density matrix of a
mixed state are given by
\[
1 \geq |\vec{n}|^{2} \geq 0 \quad \text{ \ and } \quad 1 \geq 3|\vec{n}|^{2}-2\vec{n}\cdot(\vec{n}%
\star\vec{n}) \geq 0 \quad .
\]

\noindent An arbitrary diagonal density matrix of a 3-level system
will be
\[
\rho=\frac{1}{3}(I+\sqrt{3}(n_{3}\lambda_{3}+n_{8}\lambda_{8})).
\]
In this case, the mixed-state density matrix constraints reduce to
\[
0 \leq  n_{3}^{2}+n_{8}^{2} \leq 1 \quad \text{ \ and } \quad 0 \leq 2n_{8}^{3}-6n_{3}^{2}n_{8}+3n_{3}%
^{2}+3n_{8}^{2} \leq 1 \quad .
\]
The region in the $n_{3}n_{8}$-plane where both the constraints
are satisfied is bound by an equilateral triangle with vertices at
the points
\[
(n_{3},n_{8})_{R}=(\frac{\sqrt{3}}{2},\frac{1}{2})\leftrightarrow\left(
\begin{array}
[c]{ccc}%
1 & 0 & 0\\
0 & 0 & 0\\
0 & 0 & 0
\end{array}
\right)  ,
(n_{3},n_{8})_{B}=(-\frac{\sqrt{3}}{2},\frac{1}{2})\leftrightarrow\left(
\begin{array}
[c]{ccc}%
0 & 0 & 0\\
0 & 1 & 0\\
0 & 0 & 0
\end{array}
\right)  ,
\]\[
(n_{3},n_{8})_{G}=(0,-1)\leftrightarrow.\left(
\begin{array}
[c]{ccc}%
0 & 0 & 0\\
0 & 0 & 0\\
0 & 0 & 1
\end{array}
\right)  .
\]
\begin{center}
\epsfig{figure=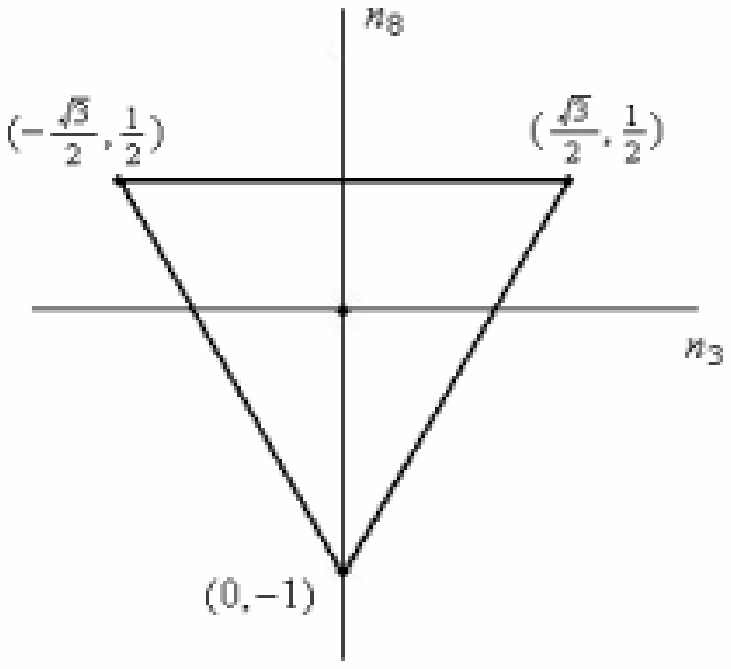}
 \end{center}
Vertices of the above triangle correspond to three mutually
orthogonal pure-states. We labeled them  Red, Blue, Green in
analogy with colored quarks [2]. In fact two pure-state vectors
$|\psi>$ and $|\psi^{\prime}>$ are orthogonal if and only if
$<\psi|\psi^{\prime}> = 0$, so that $Tr(\rho \rho^{\prime}) = 0$.
This implies $\vec{n}\cdot{\vec{n}}^{\prime} = -\frac{1}{2}$. Then
$\arccos(\vec{n}\cdot{\vec{n}}^{\prime}) = \pm\frac{2\pi}{3}$.
This is equal to the geodesic distance between two orthogonal
pure-states as measured by the standard Fubini-Study metric on
$\mathcal{C}P^{2}$.

Points on the edges of the triangle correspond to mixing of two orthogonal
pure-states of qutrits. In particular, at mid-points where bi-sectors
intersect with the edges we have
\[
(n_{3},n_{8})_{C} = ( 0,\frac{1}{2}) \leftrightarrow\frac{1}{2}\left(
\begin{array}
[c]{ccc}%
1 & 0 & 0\\
0 & 1 & 0\\
0 & 0 & 0
\end{array}
\right)  , (n_{3},n_{8})_{B} = (\frac{\sqrt{3}}{4}, -\frac{1}{4})
\leftrightarrow\frac {1}{2}\left(
\begin{array}
[c]{ccc}%
1 & 0 & 0\\
0 & 0 & 0\\
0 & 0 & 1
\end{array}
\right)  ,
\]
\[
(n_{3},n_{8})_{A} = (-\frac{\sqrt{3}}{2},-\frac{1}{4}) \leftrightarrow\frac
{1}{2}\left(
\begin{array}
[c]{ccc}%
0 & 0 & 0\\
0 & 1 & 0\\
0 & 0 & 1
\end{array}
\right)  .
\]
Triply mixed-states correspond to points inside the triangle. In particular
the origin
\[
(n_{3},n_{8})_{O} = (0,0) \leftrightarrow\frac{1}{3} \left(
\begin{array}
[c]{ccc}%
1 & 0 & 0\\
0 & 1 & 0\\
0 & 0 & 1
\end{array}
\right)
\]
corresponds to the maximally mixed state.

For 2-level systems, the orbit of any diagonal $ 2 \times 2$
density matrix of qubits
\[
\rho= \frac{1}{2}\left(
\begin{array}
[c]{cc}%
1 + n_{3} & 0\\
0 & 1-n_{3}%
\end{array}
\right)  \leftrightarrow(0,0,n_{3})
\]
under the action of the unitary group \textsf{SU(2)},
\[
\rho\rightarrow U \rho U^{\dag}%
\]
where $U \in\mathsf{SU(2)}$ (i.e. adjoint representation of
\textsf{SU(2)}, which is \textsf{SO(3)} applied on $n_{3}$) sweeps
the whole Poincar\'{e} sphere $S^{2}$. Adjoint representation
\textsf{Ad }of a given \textsf{U }$\in$ \textsf{SU(2)} is
explicitly
\[
\mathsf{Ad(U)}_{ij}\mathsf{=}\frac{1}{2}\mathsf{Tr}(\sigma_{i}\mathsf{U}%
\sigma_{j}\mathsf{U}^{\dag})\in\mathsf{SO(3)} ,
\]
so that $n_{j} \rightarrow\mathsf{Ad(U)}_{jk} n_{k} .$

In a similar way, for 3-level systems the orbit of each point
$(n_{3},n_{8})$ of the above triangle under the unitary action of
\textsf{SU(3)} (i.e. adjoint representation of \textsf{SU(3)})
will provide a generalization of the Poincar\'{e} sphere to
3-level systems. Adjoint representation \textsf{Ad }of a given
\textsf{U }$\in$ \textsf{SU(3) } is found as follows:
\[
\text{\textsf{Ad(U)}}_{ij}\mathsf{=}\frac{1}{2}\text{\textsf{Tr}}(\lambda
_{i}\mathsf{U}\lambda_{j}\mathsf{U}^{\dag})\in\text{\textsf{SO(8)}}
\]
so that
\[
n_{j} \rightarrow\text{\textsf{Ad(U)}}_{jk} n_{k} .
\]
In fact \textsf{Ad(SU(3))} is an 8-parameter subgroup of the
28-parameter rotation group \textsf{SO(8)}.

We also consider the  \textit{entropy of mixing} of $\rho$ defined
as
\[
E(\rho)= - x_{1} \log_{3} (x_{1})- x_{2} \log_{3} (x_{2})- x_{3}
\log_{3} (x_{3})  \quad .
\]
Since in diagonal form
\[
\rho= \frac{1}{3}\left(
\begin{array}
[c]{ccc}%
1+\sqrt{3}n_{3}+n_{8} &  & \\
& 1-\sqrt{3}n_{3}+n_{8} & \\
&  & 1-2n_{8}%
\end{array}
\right)  ,
\]
the entropy of mixing of $\rho$ becomes
\begin{align}
E(\rho)=  &
-(\frac{1+\sqrt{3}n_{3}+n_{8}}{3})\log_{3}(\frac{1+\sqrt{3}
n_{3}+n_{8}}{3}) - (\frac{1-\sqrt{3}n_{3}+n_{8}}{3})\log_{3}(\frac{1-\sqrt{3}n_{3}+n_{8}}%
{3})\nonumber\\
&  - (\frac{1-2n_{8}}{3})\log_{3}(\frac{1-2n_{8}}{3}) \quad .
\nonumber
\end{align}
The equi-mixing curves in the $n_{3}n_{8}$-plane are shown on the
following diagram:
\begin{center}
\epsfig{figure=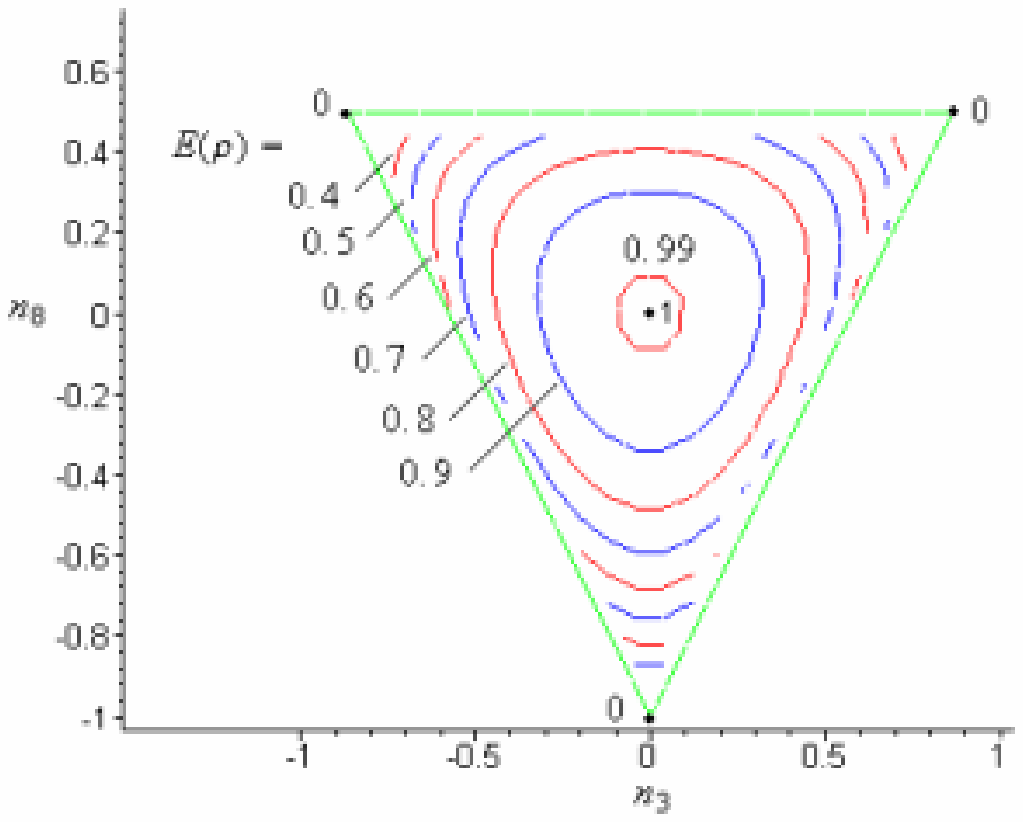}
\end{center}
\bigskip

\noindent \textbf{Acknowledgement}

\noindent We thank Professor A. Shumovsky for bringing Ref.[2] to
our attention and Professor V. Manko for comments.

\medskip

\noindent \textbf{References}
\begin{description}
 \item {\small {[1]} C. M. Caves, G. J. Milburn, Opt. Commun. \textbf{179}, 439
 (2000)}
 \item {\small {[2]} A. V. Burlankov, D.
N.Klyshko, JETP Letters \textbf{69}, 839 (1999)} \item
{\small{[3]} M. V. Chekova, L. A. Krivitsky, S. P. Kulik, G. A.
Maslennikov, Phys. Rev. A\textbf{70}, 053801 (2004)} \item {\small
{[4]} G. Khanna, S. Mukhopadhyay, R. Simon, N. Mukunda, Ann. Phys.
\textbf{253},55 (1997) }
\end{description}

\end{document}